\documentstyle[aps]{revtex}

\newcommand{\be}{\begin{equation}}
\newcommand{\ee}{\end{equation}}
\newcommand{\bea}{\begin{eqnarray}}
\newcommand{\eea}{\end{eqnarray}}
\newcommand{\bml}{\begin{mathletters}}
\newcommand{\eml}{\end{mathletters}}
\newcommand{\pa}{\partial}
\newcommand{\dxy}{\delta(\vec{x}-\vec{y})}

\newcommand{\e}{\epsilon}
\newcommand{\ve}{\varepsilon}
\newcommand{\vx}{\vec{x}}
\newcommand{\vy}{\vec{y}}
\newcommand{\vk}{\vec{k}}
\begin{document}
\draft
\title{Poincar\'e Invariance of a Quantized Duality Symmetric 
Theory\cite{byline}}
\author{H. O. Girotti}
\address{Instituto de F\'{\i}sica,
Universidade Federal do Rio Grande do Sul \\ Caixa Postal 15051, 91501-970  -
Porto Alegre, RS, Brazil.}
\maketitle

\begin{abstract}
The noncovariant duality symmetric action put forward by Schwarz-Sen is 
quantized by means of the Dirac bracket quantization procedure. The resulting
quantum theory is shown to be, nevertheless, relativistically invariant. 
\end{abstract}
\pacs{PACS: 11.30.Cp, 11.30.Ly, 11.25.-w}

\narrowtext

In Ref.\cite{Schz}, Schwarz and Sen proposed a class of gauge invariant 
actions which are also invariant under discrete duality transformations. In
particular, the duality symmetric generalization of the four dimensional
Maxwell action involves two gauge potentials 
$ A^{\mu,a}(0\le{\mu}\le{3},\,\, 1\le{a}\le{2})$ and reads\cite{f1} 

\be
\label{11}
S = - \frac{1}{2} \int d^4x \left( B^{a,i} \epsilon_{ab} E^{b,i} \,+\,
B^{a,i}B^{a,i} \right)\,\,\,,
\ee

\noindent
where

\bml
\label{12}
\bea
&& E^{a,i} = - F^{a,0i} = - \left( \partial^{0}A^{a,i} -
\partial^{i}A^{a,0}\right)\,\,\,, \label{mlett:a12} \\
&& B^{a,i} = -\frac{1}{2} \epsilon^{ijk}
F^{a}_{jk} = -\epsilon^{ijk}\partial_{j}A^{a}_{k}\,\,\,,\label{mlett:b12}
\eea
\eml

\noindent
and $1\le{i,j,k}\le{3}$. $S$ is separately invariant under the local gauge 
transformations

\bml
\label{13}
\bea
&& A^{a,0} \rightarrow A^{a,0} + \Psi^{a}\,\,\,, \label{mlett:a13} \\
&& A^{a,i} \rightarrow A^{a,i} - \partial^{i}\Lambda^{a}\,\,\,, 
\label{mlett:b13}    
\eea
\eml

\noindent
and under the discrete duality transformations

\be
\label{14}
A^{a,\mu} \rightarrow \epsilon_{ab} A^{b,\mu}.
\ee

\noindent
The use of the equations of motion,

\be
\label{15}
\epsilon^{ijk} \epsilon_{ab} \pa_{0} \pa_{j} A^{b}_{k} + \pa_{j}
\left(\pa^{j} A^{a,i} - \pa^{i} A^{a,j}\right) = 0\,\,\,,
\ee

\noindent
allows for the elimination from $S$ of one of the gauge fields, the action for
the remaining one being the conventional Maxwell action.

In terms of the gauge potentials, the Lagrangian density in (\ref{11}) reads 

\be
\label{16}
{\cal{L}} = \frac{1}{2} \e^{jki}(\pa_{j}A^{a}_{k})\e_{ab}(\pa_{0}A^{b}_{i})-
\frac{1}{2} \e^{jki}(\pa_{j}A^{a}_{k})\e_{ab}(\pa_{i}A^{b}_{0})
-\frac{1}{4} F^{a,jk} F^a_{jk}\,\,\,.
\ee

\noindent
Clearly, $\cal{L}$ is not a Lorentz scalar. Some alternatives have been
suggested to reconcile, already at the classical level, duality and Lorentz 
symmetries\cite{KoPa,To1}. In this paper we demonstrate that the quantum field 
theory arising from (\ref{16}) is, nevertheless, relativistically invariant.   

The present work can be summarized as follows. We start by presenting the 
Hamiltonian formulation of the model before 
gauge fixing. After choosing the Coulomb gauge, the theory is quantized 
by means of the Dirac bracket quantization procedure \cite{Di,Fr,Su,Gi}. The 
resulting quantum theory turns out to be local and quantum mechanically 
consistent. 
The next step consists in building a set of composite operators which will be 
shown to verify the Dirac-Schwinger algebra\cite{DiSch,Sch}. As a consequence, 
a set of charges obeying the Poincar\'e algebra exist and can inmediately be 
constructed. We prove, afterwards, that the full set of composite operators
obeying the Dirac-Schwinger algebra are the components of a second-rank
symmetric tensor. The transformation properties of the basic fields
under the Poincar\'e group are also studied and serve to demonstrate that 
the noncovariant Coulomb gauge condition is preserved under Lorentz boosts.
We end by arguing that our results can be generalized for an arbitrary
canonical gauge.

The canonical Hamiltonian ($H_c$) following from (\ref{16}) reads

\be
\label{21}
H_c = \int d^{3}x \left[ \frac{1}{2} \e^{jki}(\pa_{j}A^{a}_{k})\e_{ab}
(\pa_{i}A^{b}_{0}) + \frac{1}{4} F^{a,jk}F^a_{jk} \right]\,\,\,.
\ee

\noindent
Furthermore, the system possesses the primary constraints

\bml
\label{22}
\bea
\Omega^a_0 &\equiv& \pi^a_0 \approx 0\,\,\,, \label{mlett:a22} \\
\Omega^a_i &\equiv& \pi^a_i + \frac{1}{2} \e_{ab}\,\e_{ijk}\,\pa^j A^{b,k}
\approx 0\,\,\,,\label{mlett:b22}
\eea
\eml

\noindent
where we have designated by $\pi^a_{\mu}$ the momentum canonically conjugate to
$A^{a,\mu}$. Then, the total Hamiltonian ($H^{\prime}$) is given by
$H^{\prime} = H_c + \int d^3x \left( u^{a,0} \Omega^a_0 + 
u^{a,i} \Omega^a_i \right)$,
where the $u$'s are Lagrange multipliers. Persistence in time of $\Omega^a_0$
produces neither secondary constraints nor determines the Lagrange
multipliers. On the other hand, persistence in time of the primary constraints
$\{\Omega^a_i\}$ does not lead to the existence of secondary constraints but 
determines partially the Lagrange multipliers $\{u^a_i\}$. Indeed, since the 
Poisson bracket\cite{f3}

\be
\label{24}
[\Omega^a_i(\vec{x})\, , \, \Omega^b_j(\vec{y})]_{P} = - \e_{ab}\,\e_{ijk} 
\, \pa^j_x \dxy
\ee

\noindent
does not vanish, ${\dot{\Omega}}^a_i = [\Omega^a_i , H^{\prime}]_{P} 
\approx 0 $ yields $u^{a,i} = \e_{ab}(B^{b,i} -
\pa^i\phi^{b})$, where $\phi^{a}$ is an arbitrary scalar. Thus, 

\be
\label{25}
\Omega^{a}(\vec{x}) = \pa^i \Omega^a_i(\vec{x}) \approx 0
\ee

\noindent
and $\Omega^a_0 \approx 0 $ are the first-class constraints in the 
theory\cite{Res}. 

To isolate the second-class constraints from (\ref{mlett:b22}), we split 
$\Omega^a_i$ into longitudinal ($L$) and
transversal ($T$) components, namely, $\Omega^a_i = \Omega^a_{Li} 
+ \Omega^a_{Ti}$, where $\Omega^a_{Li} = -\frac{\pa_i \pa^j}{\nabla^2}
\Omega^a_j$, $\Omega^a_{Ti} = \left(g_i^j 
+ \frac{\pa_i \pa^j}{\nabla^2}\right)\Omega^a_j$ and 
$\nabla^2 \equiv -\pa_j \pa^j $. The first-class constraint (\ref{25}) 
only involves the longitudinal components $\Omega^a_{Li}$ and states that these
components vanish individually. Then, the second-class constraints are
 
\be
\label{28}
\Omega^a_{Ti} = \pi^a_{Ti} + \frac{1}{2} \e_{ab}\,\e_{ijk}\,\pa^j A^{b,k}_T
\approx 0\,\,\,.
\ee

The determination of the constraint structure is over. It only 
remains to be mentioned that the gauge potential $A^{a,\mu}$, when acted 
upon by the generator of infinitesimal gauge transformations, 
$G = \int d^3x \left(\Psi^a \Omega^a_0 + \Lambda^a \Omega^a\right)$,
undergoes the change $A^{a,\mu}\rightarrow A^{a,\mu}+\delta A^{a,\mu}$
with $\delta A^{a,0} = [A^{a,0},G]_{P} = \Psi^a$ and
$\delta A^{a,i} = [A^{a,i},G]_{P} = - \pa^i \Lambda^a$, in agreement with
(\ref{13}). 

We shall next quantize the model by means of the Dirac bracket quantization
procedure\cite{Di,Fr,Su,Gi}. To this end, we start by fixing the gauge 
through the subsidiary conditions

\bml
\label{29}
\bea
&&\chi^{a,0} \equiv A^{a,0} \approx 0\,\,\,,\label{mlett:a29}\\
&&\chi^a \equiv \pa_iA^{a,i} \approx 0 \,\,\,.\label{mlett:b29}
\eea
\eml

\noindent
The formulation of the quantum dynamics of a gauge theory in the Coulomb
(physical) gauge is of importance for understanding its structural aspects. 
The fact that the Coulomb condition and $A^{a,0} \approx 0$ are, when acting
together, accessible gauge conditions is a peculiarity of the model under
analysis. This is not the case, for instance, in quantum electrodynamics. 

Since the full set of constraints and
gauge conditions is, by construction, second-class, Dirac-brackets with
respect to them can be introduced in the usual manner. Afterwards,
the phase-space variables are promoted to operators obeying an equal-time
commutator algebra which is to be abstracted from the corresponding Dirac
bracket algebra, the constraints and gauge conditions thereby translating 
into strong operator relations. This is the Dirac bracket quantization 
procedure, which presently yields\cite{f4}

\bml
\label{201}
\bea
&& \bigl [ A^{a,i}_T(\vx) \,,\, A^{b,j}_T(\vy) \bigr ]\,=\,-i \e_{ab}\,\e^{ijk}
\frac{\pa^x_k}{\nabla^2}\dxy\,\,\,, \label{mlett:a201} \\
&& \bigl [ A^{a,i}_T(\vx) \,,\, \pi^b_{Tj}(\vy) \bigr ]\,=\, \frac{i}{2}
\delta_{ab}\left( g^i_j + \frac{\pa^i_x \pa_j^x}{\nabla^2} \right)\dxy \,\,\,,
\label{mlett:b201}\\ 
&& \bigl [ \pi^{a}_{Ti}(\vx) \,,\, \pi^b_{Tj}(\vy) \bigr ]\,=\, \frac{i}{4}
\e_{ab}\, \e_{ijk} \pa^k_x \dxy \,\,\,. \label{mlett:c201}  
\eea
\eml

\noindent
As for the quantum mechanical Hamiltonian ($H$), it can be read off from
(\ref{21}) after taking into account that constraints and gauge conditions
act, within the algebra (\ref{201}), as strong operator identities. Then

\be
\label{202}
H = \frac{1}{4} \int d^3x  F^{a,jk}F^a_{jk}\,=\,-\frac{1}{2}\int d^3x 
B^{a,j}B^a_j \,\,\,.
\ee

\noindent
One may wonder on whether the right hand side of (\ref{202}) is afflicted by 
ordering ambiguities. However, this not so, since

\be
\label{203}
\left[B^{a,i}(\vx)\,,\,B^{b,j}(\vy)\right]\,=\,i\,\e_{ab}\,\e^{ijk}\,
\pa_k^x\dxy \,\,\,,
\ee

\noindent
as follows from (\ref{mlett:a201}) and (\ref{mlett:b12}). 

The Hamilton equations of motion arising from (\ref{201}) and (\ref{202}) are

\bml
\label{204}
\bea
&& {\cal {D}}^{(-)ab}_{ik} A^{b,k}_T \,=\,0 \,\,\,,\label{mlett:a204}\\
&& \pa_0 \pi^a_{T i} \,=\,\frac{1}{2} \pa^j F^a_{ji}\,\,\,,\label{mlett:b204} 
\eea
\eml

\noindent
where

\be
\label{205}
{\cal {D}}^{(\pm) ab}_{ik}\equiv g_{ik} \delta_{ab}\pa_0\,
\pm \,\e_{ab}\e_{ijk}\pa^j \,\,\,.
\ee

\noindent
Notice that, in the Coulomb gauge, the Lagrange equation of motion (\ref{15})
can be casted as

\be
\label{206}
\e^{jli} \pa_l \, {\cal {D}}^{(-) ab}_{ik} A^{b,k}_T \,=\,0 \, \Longrightarrow 
 {\cal {D}}^{(-) ab}_{ik} A^{b,k}_T \,=\,\pa_i \xi^a \,\,\,.
\ee

\noindent
Since $\pa^i {\cal {D}}^{(-) ab}_{ik} A^{b,k}_T \,=\,0$, the function $\xi^a$ must
verify $\nabla^2 \xi^a = 0$ but is otherwise arbitrary. Thus, the Lagrangian 
and the Hamiltonian formulations lead to equivalent equations of motions only 
after the introduction of a regularity requirement at spatial infinity. This 
situation resembles that encountered in connection with the theory of the 
two-dimensional ($x^0, x^1, x^{\pm} = 1/\sqrt{2}(x^0 \pm x^1)$) self-dual field
($\Phi$) put forward by Floreanini and Jackiw\cite{FoJa,CoGi}, 
where the equations of motion in the Lagrangian and Hamiltonian formulations 
turn out to be, respectively, $\pa_1 \pa_{-}\Phi = 0$ and $\pa_{-}\Phi = 0$. 
We also recall that in order to solve $\pa_{-}\Phi = 0$ one starts by 
realizing that
$\pa_{-}\Phi = 0 \Longrightarrow \pa _{+} \pa_{-}\Phi = 0 \Longrightarrow 
\Box \Phi = 0$. The solutions of $\pa_{-}\Phi = 0$ are then contained in the
field of solutions of $\Box \Phi = 0$. We shall follow here a
similar approach, since

\be
\label{207}
{\cal {D}}^{(-) ab}_{ik} A^{b,k}_T \,=\,0 \Longrightarrow 
{\cal {D}}^{(+) ca,li} {\cal {D}}^{(-) ab}_{ik} A^{b,k}_T \,=\,0 
\Longrightarrow \Box A^{c,l}_T = 0\,\,\,. 
\ee        

\noindent
The solving of $\Box A^{a,i}_T = 0$ leads to

\be
\label{208}
A^{a,i}(x)\,=\,\int d^3y D(x - y){\buildrel{\scriptstyle{\leftrightarrow}}
\over{\textstyle{\partial}}}^0_y A^{a,i}_T(y)\,\,\,,
\ee

\noindent
where $D(x -y)$ is the zero-mass Pauli-Jordan delta function and
$(A{\buildrel{\scriptstyle{\leftrightarrow}}  \over{\textstyle{\partial}}
}^k B) \equiv A \partial^k B - B \partial^k A $. From this
last equation and (\ref{201}) follows that the field commutator at different
space-time points reads

\be
\label{209}
\bigl[A^{a,i}_T(x)\,,\,A^{b,j}_T(y) \bigr]\,=\,i\,\left[\delta_{ab} 
\left( g^{ij} + \frac{\pa^i_x \pa^j_x}{\nabla^2_x} \right)
\,-\,\e_{ab} \e^{ijk} \frac{\pa^x_k \pa^x_0}{\nabla^2_x} \right]\,
D(x - y)\,\,\,.
\ee

\noindent
By applying ${\cal {D}}^{(-) ca}_{ki}(x)$ to both sides of (\ref{209}), one can
check that the field configurations entering the just mentioned commutator are
in fact solutions of (\ref{mlett:a204}). As known, the function $D(x - y)$ 
can be given as the sum of a positive plus a negative frequency
part and we, therefore, can write

\be
\label{210}
A^{a,i}_T(x) \,=\,A^{a,i (+)}_T(x)\,+\,A^{a,i (-)}_T(x)\,\,\,,
\ee
  
\noindent
where

\be
\label{211}
A^{a,i (\pm)}_T(x)\,=\,\frac{1}{(2\pi)^{3/2}}\int
\frac{d^3k}{\sqrt{2|\vec{k}|}} \exp [\pm i(|\vec{k}|x^0 - \vec{k}\cdot \vx)]
\sum_{\lambda = 1}^{2}\ve^{a,i}_{\lambda}(\vec{k})a^{(\pm)}_{\lambda}(\vk)
\,\,\,
\ee

\noindent
and $\ve^{a,i}_{\lambda}(\vec{k}), \lambda = 1,2,$ are unit norm 
polarization vectors. By going back with (\ref{211}) into (\ref{209}) one 
obtains

\be
\label{212}
\sum_{\lambda,\lambda^{\prime} = 1}^{2} \ve^{a,i}_{\lambda}(\vec{k})
\ve^{b,j}_{\lambda^{\prime}}(\vec{k^{\prime}})\left[a^{(-)}_{\lambda}(\vk)\,,\,
a^{(+)}_{\lambda^{\prime}}(\vec{k^{\prime}})\right] = \left[-\delta_{ab}
\left(g^{ij} + \frac{k^i k^j}{|\vk|}\right)\,+\,\e_{ab}\e^{ijl}\frac{k_l}
{|\vk|}\right]\delta(\vk - \vec{k^{\prime}})\,\,\,,
\ee

\noindent
while all others commutators vanish. The polarization vectors are to be found
by replacing (\ref{211}) into the gauge condition (\ref{mlett:b29}) and the
equation of motion (\ref{mlett:a204}). In this way we arrive, respectively, to
$k_i \ve^{a,i}_{\lambda}(\vec{k}) = 0$ and

\be
\label{214}
\Sigma^{ab}_{ij}(\vk) \ve^{b,j}_{\lambda}\,=\,0\,\,\,,
\ee

\noindent
where 

\be
\label{215}
\Sigma^{ab}_{ij}(\vk) \equiv
g_{ij}\, \delta_{ab}\, k_0\,-\,\e_{ab}\,\e_{ilj}\,k^l\,\,\,. 
\ee

\noindent
The vanishing of the determinant of the matrix $\Sigma^{ab}_{ij}$ is a
necessary and sufficient condition for the homogeneous system of equations in
(\ref{214}) to have solution different from the trivial one
$\ve^{b,j}_{\lambda} = 0$. In the present case this determinant is proportional
to $k^2$ and its vanishing merely states that the theory only propagates
zero-mass particles. Furthermore, (\ref{214}) also implies that 
$\ve^{a,i}_{\lambda} 
\Sigma^{ab}_{ij} \ve^{b,j}_{\lambda} = 0 $. This nontrivial relationship among
the polarization vectors associated with different gauge potentials can be
casted as 

\be
\label{216}
\sum_{\lambda = 1}^{2} {\vec{\ve}}^{\,\, a}_{\lambda}(\vk)\, \times\, 
{\vec{\ve}}^{\,\, b}_{\lambda}(\vk) \,
= \, -\, 2\, \e_{ab} \,\frac{\vk}{|\vk|}\,\,\,.
\ee

\noindent
On the other hand, the Coulomb
gauge polarization vectors span, by construction, the space orthogonal to 
$\vk$, i.e.,

\be
\label{217}
\sum_{\lambda = 1}^{2} \ve^{\,\, a,i}_{\lambda}(\vk) \,
\ve^{\,\, a,j}_{\lambda}(\vk) \,
= \,-\left( g^{ij}\,+\,\frac{k^i k^j}{|\vk|^2}\right) \,\,\,.
\ee

\noindent
By using (\ref{216}) and (\ref{217}) we can solve at once for the commutator 
in (\ref{212}) and find

\be
\label{218}
\left[a^{(-)}_{\lambda}(\vk)\,,\,
a^{(+)}_{\lambda^{\prime}}(\vec{k^{\prime}})\right]\,=\,\delta_{\lambda
\lambda^{\prime}}\, \delta(\vk\,-\,\vec{k^{\prime}})\,\,\,.
\ee

\noindent
Thus the space of states is, as expected, a Fock space with positive definite 
metric. 

Hence, the quantization of the Schwarz-Sen model has led to a local and
physically sensible quantum field theory. Our next task is to demonstrate that
this quantum theory is also relativistically invariant. 

We are therefore looking for a set of composite operators 
$\{\Theta_{\mu \nu}\}$ which may serve as Poincar\'e densities. By
experience, we try to build them according to the following rules 

\be
\label{301}
\Theta_{\mu \nu}\,=\,T_{\mu \nu}\,+\,\pa^{\lambda}\psi_{{\underline{\lambda
\mu}} \nu}\,\,\,,
\ee

\noindent
where

\be
\label{302}
T_{\mu \nu}\,=\,\frac{\pa {\cal{L}}}{\pa (\pa^{\mu} A^{a, \rho})}\,
\pa_{\nu}A^{a, \rho}\,-\,g_{\mu \nu}\,\cal{L}\,\,\,,
\ee

\be
\label{303}
\psi_{{\underline{\lambda \mu}} \nu}\,=\,\frac{1}{2}\left(
S_{{\underline{\lambda \mu}} \nu}\,+\,S_{{\underline{\nu \mu}} \lambda}\,
+\,S_{{\underline{\lambda \nu}} \mu} \right)\,\,\,,
\ee

\be
\label{304}
S_{{\underline{\lambda \mu}} \nu}\,=\,-\,
\frac{\pa {\cal{L}}}{\pa (\pa^{\nu} A^{a, \alpha})}
{\cal{A}}^{\alpha}_{{\underline{\lambda \mu}} \beta}\,A^{a,\beta}\,\,\,,
\ee

\noindent
and ${\cal{A}}^{\alpha}_{{\underline{\lambda \mu}} \beta} =
g^{\alpha}_{\lambda} g_{\mu \beta}\,
-\,g^{\alpha}_{\mu} g_{\lambda \beta}$. Clearly, $\psi$, $S$ and ${\cal{A}}$ 
are antisymmetric under the exchange of the underlined indices.
These are, of course, the standard rules for constructing the symmetric 
(Belinfante) 
energy-momentum tensor. However, we can not yet decide on whether 
or not $ \Theta $ is a tensor\cite{f5}. 
By bringing (\ref{16}) into (\ref{302}) and (\ref{304}) one obtains

\bml
\label{305}
\bea
\Theta_{00}\,&=&\,-
\frac{1}{2}\,B^{a,i}\,B^{a}_{i}\,\,\,,\label{mlett:a305} \\
\Theta_{0i}\,&=&\,\Theta_{i0}\,=\,-
\frac{1}{2}\,\e_{ijk}\,\e_{ab} B^{a,j}\,B^{b,k}\,\,\,,\label{mlett:b305} \\
\Theta_{ij}\,&=&\,\Theta_{ji}\,=\,-  B^{a}_{i}\,B^{a}_{j}\,+\,g_{ij}\,
 B^{a,l}\,B^{a}_{l}\,\,\,.
\eea
\eml

\noindent
Thus, $\Theta$ is symmetric and free of ordering ambiguities.
  
We look next for the equal-time commutator algebra obeyed by the components of 
$\Theta $. According to (\ref{305}), this algebra is fully determined
by the commutator (\ref{203}). In particular, one can 
corroborate that 

\bml
\label{306}
\bea
\left[\,\Theta^{00}(x^0,\vec{x})\,,\,\Theta^{00}(x^0, \vec{y})\,\right]
&=& -\,i\,\left\{\,\Theta^{0k}(x^0,\vec{x})\,
+\,\Theta^{0k}(x^0, \vec{y})\,\right\} \,\partial^x_k\delta(\vec{x}-\vec{y})\,
\,\,,\label{mlett:a306} \\
\left[\,\Theta^{00}(x^0,\vec{x})\,,\,\Theta^{0k}(x^0, \vec{y})\,\right]
&=& \,-\,i\,\left\{\,\Theta^{kj}(x^0,\vec{x})\,-\,g^{kj}\,\Theta^{00}(x^0, \vec{y})
\,\right\}\, \partial^x_j \,{\delta(\vec{x}-\vec{y})}
\,\,\,,\label{mlett:b306} \\
\left[\,\Theta^{0k}(x^0,\vec{x})\,,\,\Theta^{0j}(x^0, \vec{y})\,\right]
&=& i\,\left\{\,\Theta^{0k}(x^0, \vec{y})\,\partial^j_x\,
+\,\Theta^{0j}(x^0,\vec{x})\,\partial^k_x\,\right\}\,{\delta(\vec{x}-\vec{y})}
\,\,\,, \label{mlett:c306}
\eea
\eml

\noindent
which is just the Dirac-Schwinger algebra\cite{DiSch}. As it is well 
known\cite{DiSch}, this guarantees that the charges 

\bml
\label{307}
\bea
&& P^{\mu}\,\equiv\,\int d^3x\,\Theta^{0 \mu}\,\,\,, \label{mlett:a307}\\
&& J^{\mu \nu}\,\equiv\,\int d^3x \left(\Theta^{0 \mu} x^{\nu}\,-\,
\Theta^{0 \nu} x^{\mu}\right)\,\,\,,\label{mlett:b307}
\eea
\eml

\noindent
obey the Poincar\'e algebra, i.e.,
$\left[P^{\mu} , P^{\nu}\right] = 0$,
$\left[J^{\mu \nu} , P^{\sigma}\right] = i \left(g^{\mu \sigma} 
P^{\nu}\,-\,g^{\nu \sigma} P^{\mu}\right)$ and
$\left[J^{\mu \nu} , J^{\rho \sigma}\right] = i
\left( g^{\mu \rho}J^{\nu \sigma} + g^{\nu \sigma}J^{\mu \rho} -
g^{\mu \sigma}J^{\nu \rho} - g^{\nu \rho}J^{\mu \sigma}\right)$.

It takes just a few more steps to demonstrate that $\Theta $ is a
tensor.  Indeed, the additional equal-time commutators 
$\left[\,\Theta^{ij}(x^0,\vec{x})\,,\,\Theta^{00}(x^0, \vec{y})\,\right]$ and
$\left[\,\Theta^{ij}(x^0,\vec{x})\,,\,\Theta^{0k}(x^0, \vec{y})\,\right]$ can
also be readily evaluated by using (\ref{305}) and (\ref{203}). These results 
and (\ref{306}) can be collected into

\bml
\label{309}
\bea
&& \left[P^{\mu}\,,\,\Theta^{\alpha \beta}\right] =
-i\,\pa^{\mu}\Theta^{\alpha \beta}\,\,\,,\label{mlett:a309}\\
&&\left[J^{\mu \nu}\,,\,\Theta^{\alpha \beta}\right] =
-i\left(x^{\nu}\pa^{\mu} - x^{\mu}\pa^{\nu}\right)\Theta^{\alpha \beta}
-i\left(\Theta^{\mu \alpha}g^{\nu \beta} +
\Theta^{\mu \beta}g^{\nu \alpha} - \Theta^{\nu \alpha}g^{\mu \beta} -
\Theta^{\nu \beta}g^{\mu \alpha}\right)\,\,\,,\label{mlett:b309}
\eea
\eml

\noindent
which are, respectively, the translation and rotation transformation laws to be
obeyed by a second-rank tensor\cite{Sch1}. The purported proof of relativistic
invariance of the quantized Schwarz-Sen theory is now complete. 

What remains to be done is to demonstrate that the Coulomb gauge formulation of
the quantized Schwarz-Sen theory is in fact covariant. Since translations and 
ordinary rotations do not destroy the Coulomb gauge condition we concentrate 
on Lorentz boosts. By using (\ref{mlett:b307}), (\ref{305}), (\ref{mlett:b12})
and (\ref{mlett:a201}) one finds that

\be
\label{310}
- i \left[J^{0k}\,,\,A^{a,i}_{T}\right]\,=\,(x^0\,\pa^{k}
\,-\,x^{k}\,\pa^{0})A^{a,i}_{T}\,-\,
\e_{ab}\e^{klj} \frac{\pa^{i}\pa_{l}}{\nabla^2} A^{b}_{T j}\,\,\,.
\ee

\noindent
The term proportional to $\e_{ab}$ signalizes that gauge potentials 
corresponding to different values
of $a$ get mixed by Lorentz boosts. This does not occur for ordinary 
rotations. Furthermore, the mixing term in (\ref{310}) describes an operator 
gauge transformation, which, as one easily verifies, makes 
this commutator compatible with the transversality condition $\pa_i A^{a,i}_T 
= 0$. Hence, under Lorentz boosts, the field $A^{a,i}_{T}$ undergoes, besides 
the usual vector transformation, an operator gauge transformation 
which restores the Coulomb gauge in the new Lorentz frame\cite{Zu}.
  
Although this work has been entirely carried out within the Coulomb gauge, we 
observe that the quantized Schwarz-Sen model turned out to be a local theory 
fully formulated in terms of the gauge invariant fields 
$A^{a,j}_T, a = 1, 2$. Therefore, our conclusions about relativistic 
invariance apply equally well for all canonical gauges.


\begin{references}
\bibitem {f1} Our space-time metric is $g_{00}=-g_{11}=-g_{22}=1$ while 
$\epsilon_{ab}$ designates a generic element of the two-dimensional 
antisymmetric unit matrix ($\epsilon_{12}=+1$).
\bibitem {f3} The time label in the arguments of the fields will be
omitted whenever it does not lead to confusion.
\bibitem {f4} We shall not distinguish between a quantum field operator and 
its classical counterpart.
\bibitem {f5} Recall that ${\cal{L}}$ is not Lorentz scalar and that 
a set of four quantities $A^{a,\mu}, \mu = 0,1,2,3$,
obeying the Coulomb condition do not transform as components of a four-vector 
under Lorentz boosts.
\end{references}
\end{document}